\documentclass[,final]{aipproc}
\layoutstyle{6x9}

\def\sqr#1#2{{\vcenter{\hrule height.#2pt\hbox{\vrule width.#2pt
height#1pt \kern#1pt \vrule width.#2pt}\hrule height.#2pt}}}
\def\square{\mathchoice\sqr64\sqr64\sqr{4.2}3\sqr{3.0}3}

\begin{document}

\def\negenspace{\kern-1.1em}
\def\quer{\negenspace\nearrow}

\title{Anomalies and gravity}
\classification{PACS no.: 04.50.+h; 04.20.Jb; 03.50.Kk}
\keywords{Gauge theories, anomalies, gravity, torsion, Pontrjagin
terms}
\author{Eckehard W. Mielke\thanks{E-mail: ekke@xanum.uam.mx}}{
address={
Universidad Aut\'onoma Metropolitana--Iztapalapa,\\
Apartado Postal 55-534, C.P. 09340, M\'exico, D.F., MEXICO}}

\maketitle
\begin{abstract}
Anomalies in Yang-Mills type gauge theories of gravity are reviewed.
Particular attention is paid to the relation between the Dirac spin,
the axial current $j_5$ and the non-covariant gauge spin $C$.  Using
diagrammatic techniques, we show that only generalizations of the
$U(1)$- Pontrjagin  four--form $F\wedge F= dC$ arise  in the {\em
chiral anomaly}, even when coupled to gravity. Implications for
Ashtekar's canonical approach to quantum gravity are discussed.
\end{abstract}

\section{Introduction: Anomalies for pedestrians}
Anomalies can be viewed as a breaking of some Noether symmetry
through the effects of the vacuum.  In relativistic quantum field
theory (QFT), such a (classical) symmetry  is {\em broken} by field
quantization, cf. Refs \cite{Holst,Be00,VH05} for recent reviews.
This has important implications on such physical processes as the
decay of the neutral $\pi$-meson \cite{BJ69}, induced instanton
effects \cite{NWK02}, or underlies the postulation of the {\em
axion} in quantum chromodynamics (QCD), cf. Ref. \cite{MR06}.

In  quantum electrodynamics (QED),  Schwinger \cite{Sc51}
demonstrated that the charge current $j$ can be retained conserved,
i.e. $\langle dj \rangle =0$, whereas the conservation of the axial
current $j_5$ is broken, $\langle dj_5\rangle \neq 0$.

In the Fujikawa approach \cite{Fu79}, the right-hand side  can
obtain from considering the point-splitted current
$j_5(x;\epsilon):=\overline{\psi}(x)\gamma_5{}^*\gamma\psi(x+\epsilon)$,
where $\epsilon$ is an infinitesimal four-vector in spacetime. Such
an expression can be rendered invariant by dressing it with a
path-ordered exponential
\begin{equation}
\overline{\psi}(x)\gamma_5{}^*\gamma\psi(x+\epsilon)\rightarrow
\overline{\psi}(x)\gamma_5{}^*\gamma\psi(x+\epsilon){\bf P}
\exp\left\{i\int_x^{x+\epsilon}  A\right\}\, .
\end{equation}
The variation $\delta/\delta  A$ of the current $j_5(x;\epsilon)$ is
compensated by the variation of the exponential. As the parallel
transport from $x^i \to x^i+\epsilon^i$ along the infinitesimal line
element can be expanded perturbatively, it is clear that the net
effect of this approach is just the standard result $\langle
dj_5(x)\rangle=2im \langle P \rangle-(1/96\pi^2) F\wedge F$ for
massive fermions, where $F:=dA$ is the gauge field strength. Further
details of the path integral formulation were developed, e.g., in
Refs. \cite{AUV88,AUV89,UV92}  with extension of the regularized
Jacobian, as well as in the light-cone gauge \cite{GSV96} of the
Schwinger model.

There is an intuitive physical interpretation of this result: The
additional Chern-Simons (CS) term $C:=A\wedge dA$ corresponds to the
spin or helicity of the photon, with its spacelike part $\vec A
\cdot \vec B$ known as  magnetic helicity \cite{JP00}. Since the
axial current $j_5$ is proportional to the spin of a fermion, the
deformed current $ \widetilde j_5:= j_5 +(1/96\pi^2) A\wedge d A$,
includes the spin of the photon, lacking, however, gauge invariance.
The chiral anomaly can then be understood as the `conservation law'
\begin{equation}
\langle d\widetilde j_5\rangle =0\, ,
\end{equation}
 such that in QFT ``...the flow
of electronic spin drags some photon spin and vice versa"
\cite{WS88}.

Anomalies were studied also in Yang-Mills type gauge models of
gravity \cite{PRs} with Einsteinian instanton solutions \cite{MR05}.
Then, the equivalence principle not only requieres a coupling of
gravity to the energy-momentum current of matter, but also to the
spin current. Here we will focus on the intricate interaction
between the {\em chiral anomaly} and the spin or helicity of the
gravitational gauge field and extend it to post-Riemannian
spacetimes with torsion.
\section{Dirac fields in Riemann--Cartan spacetime}
In our notation \cite{Mi01}, a Dirac field is a bispinor--valued
zero--form $\psi$ for which $\overline{\psi}:=\psi^\dagger\gamma_0$
denotes the Dirac adjoint. The minimal coupling to the gauge
(electromagnetic) potential $A=A_i dx^i$ is accounted for via ${\cal
D}:= D + iA\wedge$, where $D\psi:= d\psi + \Gamma\wedge \psi$ is the
exterior covariant derivative with respect to the Riemann-Cartan
(RC) connection one-form $\Gamma^{\alpha\beta} =\Gamma_i
{}^{\alpha\beta}dx^i$.

The Dirac Lagrangian is given by the manifestly
 {\em Hermitian} four--form
\begin{equation}
 L_{\rm D}=L(\gamma,\psi,{\cal D}\psi)=
  {i\over 2}\left\{\overline{\psi}\,{^*\gamma}\wedge {\cal D}\psi
  +\overline{{\cal D}\psi}\wedge{^*\gamma}\,\psi\right\}+m\,
\overline{\psi}\psi\eta\,,
\label{eq:ldirac}
\end{equation}
where $\gamma:= \gamma_\alpha \vartheta^\alpha$
is the Clifford algebra-valued
coframe, see the Appendix.

The Dirac equation and its adjoint can be  obtained by varying $
L_{\rm D}$ independently with respect to $\overline{\psi}$ and
$\psi$. Making use of the torsion $\Theta:= D\gamma$ and of the
properties of the Hodge dual, the Dirac equation assumes the form
\begin{equation}
i \,^*\gamma\wedge \left({\cal D}  +{i\over 4}\,m\gamma-{1\over 2}\,
T\right)\psi=0\,, \label{di3}
\end{equation}
where
$T:= {1\over 4}\, Tr\left(\check{\gamma} \rfloor\Theta\right)
=e_\alpha\rfloor T^\alpha$
is the one--form of the trace (or vector) torsion. However, the covariant
derivative $D$ also contains torsion.

In order to separate out the purely Riemannian piece from torsion terms,
let us decompose the Riemann--Cartan connection $\Gamma=\Gamma^{\{\}}-K$
into the Riemannian (or Christoffel) connection
$\Gamma^{\{\}}$ and the {\em contortion} one--form
$K= {i\over 4} K^{\alpha\beta}\,\sigma_{\alpha\beta}$, obeying
$D\gamma= [\gamma, K]=\gamma_\alpha T^\alpha$.
Accordingly, the Dirac
Lagrangian (\ref{eq:ldirac}) splits \cite{Mi04} into a Riemannian and a
spin--contortion piece:
\begin{eqnarray}
L_{\rm D} &=& L(\gamma,\psi,D^{\{\}}\psi)-{i\over 2}
\overline{\psi}\left({^*\gamma}\wedge K - K\wedge
  {^*\gamma}\right)\psi + A\wedge j\nonumber \\
&=&L(\gamma,\psi,D^{\{\}}\psi)
  +{1\over 4}\,{\cal A}  \wedge j_5 + A\wedge j \nonumber \\
  &=&L(\gamma,\psi,D^{\{\}}\psi)  - T^\alpha \wedge \mu_\alpha + A\wedge j\, .
\label{decldirac}
\end{eqnarray}
The covariant derivative with respect to the Riemannian connection  $\Gamma^{\{\}}$
satisfies $D^{\{\}}\gamma=0$.
Hence, in a RC spacetime a Dirac spinor
only feels the {\em axial torsion} one--form
\begin{equation}
{\mathcal A}:= \frac{1}{4}  \,^* Tr(\gamma\wedge
D\gamma)= \,^*(\vartheta^\alpha\wedge T_\alpha)=
\frac{1}{2}T^{[\alpha\beta\gamma]}\eta_{\alpha\beta\gamma} =
{\cal A}_i\, dx^i \, ,  \label{axtorsion}
\end{equation}
which is invariant under  Weyl rescalings
and  {\em chiral transformations}
$\gamma \rightarrow  \gamma^\beta= e^{i\gamma^5\beta} \gamma
e^{-i\gamma^5\beta}$ of the coframe, but odd under
parity $P:\vartheta^B \rightarrow  -\vartheta^B $, where $B=1,2,3$, cf.
Ref. \cite{MMN99}.

\section{Classical axial anomaly and spin}
Similarly as in QED, the gravitational coupled Dirac Lagrangian $
L_{\rm D}=\overline{L}_{\rm D}=L_{\rm D}^\dagger$ is {\em Hermitian}
as required, even in an anholonomic
 frame. Then  minimal coupling  prescribes us automatically
with the following {\em charge and axial currents},
respectively,
\begin{equation}
j= \overline{\psi}\, ^*\gamma \psi=j^\mu\,\eta_\mu\, , \qquad
j_5 :=
\overline{\psi}\gamma_5\,^*\gamma \psi=
{1\over 3} \overline{\psi}\sigma\wedge\gamma \psi =j_5^\mu\,\eta_\mu
\, .  \label{eq:axial}
\end{equation}

{}From the Dirac equation (\ref{di3}) and its adjoint one can
readily deduce that $dj\simeq 0$  `on shell',
 whereas for the axial current we find
the well--known ``classical axial anomaly"
\begin{equation}
dj_5 =2imP  =2im \overline{\psi}\gamma_5\psi \eta
\end{equation}
for {\em massive} Dirac fields \cite{IZ80}. The same holds in a RC
spacetime. If we restore chiral symmetry in the limit $m\rightarrow
0$, this leads to classical conservation law $dj_5 =0$ of the axial
current for massless Weyl spinors, or since $dj \simeq 0$,
equivalently, for the {\em chiral currrent} $j_\pm :={1\over 2}
\overline{\psi}(1 \pm\gamma_5)\,^*\gamma\psi = \overline{\psi}_{\rm
L,R}\,^*\gamma\psi_{\rm L,R}$.

As mentioned in the Introduction, the axial current has an intriguing relation
to the  (dynamical) {\em spin current} of the Dirac field
canonically defined  by the Hermitian three--form
\begin{eqnarray}
\tau_{\alpha\beta}&:=&{\partial L_{\rm D}\over\partial\Gamma^{\alpha\beta}}
={1\over 8}\overline{\psi}\left(\,^* \gamma\sigma_{\alpha\beta}+
\sigma_{\alpha\beta}\,^* \gamma\right)\psi \nonumber\\
&=& \frac{1}{4}\,\eta_{\alpha\beta\gamma\delta}\,\overline{\psi}
\gamma^{\delta}\gamma_5\psi\eta^{\gamma}=
\tau_{\alpha\beta\gamma}\eta^{\gamma}=
\vartheta_{[\alpha} \wedge \mu_{\beta]}\, . \label{eq:spingamma}
\end{eqnarray}
Its
 components
$\tau_{\alpha\beta\gamma}=\tau_{[\alpha\beta\gamma]}$ are
{\em totally antisymmetric}.
Equivalently, in Eq. (\ref{decldirac})
torsion merely couples to the two-form
\begin{equation}
\mu_\alpha= \frac{1}{4}\vartheta_\alpha\wedge\,^* j_5 \, ,  \label{spinpot}
\end{equation}
commonly referred to as the {\em spin-energy potential}
\cite{HMMO99,Mi04}. Consquently, we obtain the remarkable result
that he vector one-form
\begin{equation}
\mu:= e_\alpha\rfloor\mu^\alpha= \frac{3}{4}\,^* j_5 \, ,  \label{Dspin}
\end{equation}
of the {\em Dirac spin} is dual to the axial current $j_5$.

There is also a relation to the axion: In Ref. \cite{MR06} it is
tentavively assumed that the dimensionless pseudo-scalar $\theta$
serves as a potential for the axial torsion via ${\cal A}=
2d\theta$. Then,
 there arises in (\ref{decldirac}) a derivative coupling of the
 would-be axion $a=\theta f_a$
to two fermions via the CPT-invariant term
 \begin{equation}
 L_{a\psi\psi}= \frac{1}{2}  d\theta \wedge  j_5
  =\frac{1}{2f_a}  da \wedge\overline{\psi}\,^*\gamma \gamma_5\psi\, ,
  \label{twoferm}
 \end{equation}
exactly  as in the usual formulation, where
 the axial current $j_5$  is  the Noether current
associated with a spontaneously broken Peccei-Quinn symmetry $U(1)_{\rm PQ}$.

\section{Axial current in the Einstein--Cartan theory}
The Einstein--Cartan (EC) theory of a gravitationally coupled  spin
$1/2$ Dirac field provides a  {\em dynamical}  understanding of the
axial anomaly on a semi-classical  level:
 The Lagrangian reads:
\begin{equation}
L ={i\over 2\ell^2} \, Tr\left(\Omega\wedge\,^*\sigma\right)+ L_{\rm D}
={1\over 2\ell^2}\,R^{\alpha\beta}\wedge\eta_{\alpha\beta} + L_{\rm D}\, ,
\end{equation}
where $\eta^{\alpha\beta}:={}^\ast(\vartheta^\alpha\wedge
\vartheta^\beta)$ is  dual to the unit two--form.

In EC--theory, Cartan's algebraic  relation between torsion and spin
implies the following relation \cite{MMM96} between the {\em axial
current} $j_5$ of the Dirac field and the translational Chern-Simons
(CS) term (\ref{CTT}), or equivalent, for the  axial torsion
one-form:
\begin{equation}
C_{\rm TT} \cong
\frac{1}{4}\,j_5 \,, \qquad
{\cal A}=2\ell^2\,^* C_{\rm TT}=
(\ell^2/2) \overline{\psi}\gamma_5\gamma\psi\, .
\label{eq:shell}
\end{equation}

Thus  in EC-theory,  the net axial current production
\begin{equation}
dj_5 \cong 4 dC_{\rm TT}
={2\over{\ell^2}} \left(T^\alpha\wedge
T_\alpha+R_{\alpha\beta}\wedge\vartheta^\alpha\wedge\vartheta^\beta\right)
\label{eq:classan}
\end{equation}
establishes  a link to the NY four form \cite{NY} for  {\em massive}
fields.

This result, cf. \cite{mie86,MMM96}, holds on the level of  first
quantization. Since the Hamiltonian of the semi-classical Dirac
field is not bounded from below, one has to go over to second
quantization, where the vacuum expectation value $\langle
dj_5\rangle$ of the axial current picks up anomalous terms.

Restoring chiral invariance for the Dirac fields, the limit
$m\rightarrow 0$, implies that the NY four--form tends to zero ``on
shell", i.e. $dC_{\rm TT}\cong (1/4) dj_5 \rightarrow 0$. This is
consistent with the fact that a Weyl spinor does not couple to
torsion at all, because then the  axial torsion ${\cal A}$ becomes a
{\em lightlike} covector, i.e. ${\cal A}_\alpha {\cal A}^\alpha\eta
={\cal A}\wedge\,^* {\cal A} \cong (\ell^4/4)\,^*j_5\wedge j_5 =0$.
Here we implicitly assume that the light-cone structure
 of the axial covector $\,^*j_5$ is not spoiled by quantum corrections, i.e. that no
 ``Lorentz anomaly" occurs as in $n=4k +2$ dimensions \cite{Leut}.


\section{Chiral anomaly in quantum field theory}
Let us recall a couple of distinguished features of the axial
anomaly: Most prominent is its relation with the Atiyah--Singer
index theorem \cite{At96}. But also from the viewpoint of
perturbative QFT, the chiral anomaly has some features which signal
its conceptual importance. For all topological field theories and
topological effects like the anomaly, there is the remarkable fact
that it does not renormalize --- higher order loop corrections do
not alter its one-loop value. This very fact guarantees that the
anomaly can be given a topological interpretation. For the anomaly,
this is the Adler--Bardeen theorem \cite{AB69}, while other
topological field theories are carefully designed  to have vanishing
beta functions, for example. Another feature is its finiteness: in
any approach, the chiral anomaly as a topological invariant is a
finite quantity.

Now, to approach the anomaly in the context of spacetime with
torsion, let us first switch off the Riemannian curvature and
concentrate on the last but one term in the decomposed Dirac
Lagrangian (\ref{decldirac}).

Then,  this term  can be regarded  as an {\em external} axial
covector ${\cal A}$  coupled to the axial current $j_5$ of the Dirac
field in an {\em initially flat} spacetime. By applying the result
(11--225) of Itzykson and Zuber \cite{IZ80}, we find that only the
term $d{\cal A}\wedge d{\cal A}$ arises in the axial anomaly, but
{\em not} the NY type term $d\,^*{\cal A}\sim dC_{\rm TT}$ as was
recently claimed \cite{ChandiaZ}. After switching on the Yang-Mills
field $G$ as well as the curved spacetime of Riemannian geometry, we
finally obtain for the vacuum expectation value of the  {\em axial
anomaly}
\begin{equation}
 \langle dj_5\rangle= 2i m \langle\overline{\psi}\gamma_5\psi\rangle\eta -{1\over
4\pi^2} Tr(G\wedge G) - {1\over
96\pi^2}\left[2R_{\alpha\beta}^{\{\}}\wedge R^{\{\}\alpha\beta} +
d{\cal A}\wedge d{\cal A}\right]. \label{axano}
\end{equation}

This result \cite{KM01} is based on diagrammatic techniques and the
Pauli--Villars regularization scheme. In this respect, it is a
typical perturbative result, and  in agreement with
\cite{Gren86,Yajima,Wies96}  no NY term arises in the anomaly. Thus
only the Weyl invariant term $d{\cal A}\wedge d{\cal A}= -2 \vec
{\cal E} \cdot \vec {\cal B}\eta$ for the axial torsion contributes
to  the axial anomaly, resembling the $U(1)$ part $F\wedge F=
dA\wedge dA$ of the Pontrjagin term (\ref{Pontr}). Torsion terms
like $d{\cal A}\wedge d{\cal A}$ and $d\,^*{\cal A}\wedge
\,^*(d\,^*{\cal A})= 4\ell^4 V_{\rm NY}\wedge \,^*V_{\rm NY}$ have
been considered previously, as part of the
 Lagrangian, in order to
make the axial torsion propagating. Due to the geometric identity
(\ref{eq:NY}) for the NY  term $d\,^*{\cal A}=2\ell^2 dC_{\rm TT}=
2\ell^2 V_{\rm NY}$, the second term is really quartic in torsion
and not scale invariant.

A rescaling of the  tetrad has been proposed, however, one should
not ignore the presence of renormalization conditions and the
generation of a scale upon renormalization. Rescaling the tetrad
would ultimately change the wave function renormalization $Z$-factor
which would creep into the definition of the NY term,  in sharp
contrast to proper topological invariants at the quantum level,
which remain unchanged under renormalization.

With no renormalization condition available for the NY term, and
other methods obtaining it as zero, we can only conclude that the
response function of QFT to a gauge variation (this is the anomaly)
delivers no NY term. Or, saying it differently, its finite value is
zero after renormalization.

\subsection{Chiral anomaly in SUGRA}
Simple supergravity consists in a consistent coupling of the EC to the
 Rarita--Schwinger spinor-valued one-form $\Psi=\Psi_i dx^i $, cf. Refs.
 \cite{UV91,MM99} for more details.

The anomaly  for the corresponding {\em axial current} $J_5 :=
i\overline{\Psi}\wedge \gamma\wedge\Psi$ is $-21\times$ the anomaly
for Dirac fields, whereas for the corresponding supersymmetric
Yang--Mills anomaly one finds $3\times$ the Dirac result:

$$\vbox{\offinterlineskip
\hrule
\halign{&\vrule#&\strut\quad\hfil#\quad\hfil\cr
height2pt&\omit&&\omit&&\omit&\cr
& {\bf Spin }  &&{\bf Gravitational} && {\bf YM anomaly}&\cr
height2pt&\omit&&\omit&&\omit&\cr
\noalign{\hrule}
height2pt&\omit&&\omit&\cr
& 1/2&& 1 && 1  &\cr
& 3/2&& $-21$ && 3  &\cr
height2pt&\omit&&\omit&&\omit&\cr}
\hrule}$$

Depending on the asymptotic helicity states, there occur
contributions of topological origin of the Riemannian Pontrjagin or
Euler type, respectively. The role of spinors for the index theorem
and in the $4D$ Donaldson invariants via Seiberg--Witten equation
has recently been reviewed by Atiyah \cite{At96}. Six dimensional
supergravity free of gauge and gravitational anomalies is studied in
Ref. \cite{Er94}.

\section{Comparison with the heat kernel method}
In the heat kernel approach, there exists for small $t\rightarrow +0$
the asymptotic expansion
\begin{equation}
K(t, x,D\quer^2)=(4\pi)^{-n/2}\sum_{k=0}^\infty t^{(k-n)/2}
K_k(x,D\quer^2) \label{exp}
\end{equation}
of the
kernel in $n$ dimensions,
where the usual Feynman ``dagger" convention
$\, A\quer:=\check{\gamma} \rfloor A= \gamma^\alpha e_\alpha\rfloor A=
(-1)^{s+1}
\,^* \left[\,^*\gamma\wedge A\right]$ for one--forms is used.

The squared Dirac operator
\begin{eqnarray}
D\quer^2 &=& -{1\over 2} \gamma^\alpha \gamma^\beta
\left(\{D^{\{\}}_\alpha\, , D^{\{\}}_\beta\} +
[D^{\{\}}_\alpha\, , D^{\{\}}_\beta]\right) -
2im D\quer^{\{\}}\nonumber\\
 &-& {i\over 4}\gamma_5 (D\quer^{\{\}} {\cal A}\quer) +{1\over 2}\gamma_5
 \sigma^{\alpha\beta} {\cal A}_\alpha D^{\{\}}_\beta +m^2 -{1\over 2}m \gamma_5 {\cal A}\quer -
 {1\over 16} {\cal A}\quer {\cal A}\quer \nonumber\\
&\cong&- \square - {1\over 8}\sigma^{\alpha\beta}
R^{\{\}}_{\alpha\beta\mu\nu}\sigma^{\mu\nu}\nonumber\\
 &-&{i\over
4}\gamma_5 (D\quer^{\{\}} {\cal A}\quer) +{1\over 2}\gamma_5
 \sigma^{\alpha\beta} {\cal A}_\alpha D^{\{\}}_\beta -
  {1\over 16}{\cal A}_\alpha {\cal A}^\alpha - m^2\, ,
\label{sqdiop}
\end{eqnarray}
 has been explicitly calculated in
Refs.\cite{Yajima,OM97}, and the terms additional to the generally
covariant Riemannian d'Alembertian operator $\square :=
\partial_\mu \left (\sqrt{\mid g\mid } g^{\mu \nu }
 \partial_\nu \right )/\sqrt {\mid g\mid }$ are identified \cite{MK98}.
Not unexpectedly, besides the familiar Riemannian curvature
scalar, only the axial torsion (\ref{axtorsion})
  contributes to the squared
Dirac operator for {\em massive} spinor fields.

The coefficients $K_k(x, D\quer^2)$
are completely determined by the form of the
second-order differential operator $D\quer^2$, which is positive for
 Euclidean signature ${\rm diag}\; o_{\alpha\beta}=(-1, \cdots, -1)$. For
 odd $k=1,3,\dots$ these
coefficients are zero, while the first nontrivial terms
\cite{Yajima}, which potentially
could  contribute to the axial anomaly,  read
\begin{eqnarray}
 Tr(\gamma_5 K_2) &=& -d\,^* {\cal A}\, ,  \nonumber\\
 Tr(\gamma_5 K_4) &=& {1\over 6}
\left[Tr\left(R^{\{\}}\wedge  R^{\{\}}\right) - {1\over 4} d{\cal
A}\wedge d{\cal A} +d{\cal K}\right],
 \end{eqnarray}
where the higher order term $d{\cal K}=
d\,^*\!\stackrel{\smile}{D}\wedge^*\!\stackrel{\smile}{D}{}^*\!{\cal
A} $ involves the  covariant derivative $\stackrel{\smile}{D}=
D^{\{\}} +i {\cal A}\gamma_5/4$ modified by the axial torsion.

However, there is an essential difference in the physical
dimensionality of the terms $K_2$ and $K_4$. Whereas in $n=4$
dimensions the Pontrjagin type term $K_4$ is dimensionless and thus,
for $k=4$, multiplied by $t^{(k-4)/2}=1$,  the term $K_2\sim d\,^*
{\cal A} = 2\ell^2 dC_{\rm TT}$ carries dimensions. Since a {\em
massive} Dirac spinor has canonical dimension $[length]^{-3/2}$, it
scales as $\psi \sim m^{3/2}$. Moreover, the term $t=1/M^2$ is
related to the regulator mass $M\rightarrow \infty$ in Fujikawa
method \cite{Fu79}. Then the second order term in the heat kernel
expansion scales as $ - K_2/t=(2\ell^2/ t)dC_{\rm TT}\cong
(\ell^2/2t)dj_5 = (im\ell^2/t)\overline{\psi}\gamma_5\psi \sim
\ell^2 M^2 m^4\rightarrow 0$. If we assume in the renormalization
proceedure, that the fundamental length $\ell$
 does not scale (no running coupling constant), the second order term in the
 heat kernel expansion will tend to zero in
the chiral limit $m\rightarrow 0$.
In the case $m\neq 0$, this
term diverges and the Fujikawa regulator method $M\rightarrow \infty$
cannot be applied.
To rescale the coframe by $\vartheta^\alpha \rightarrow
\widetilde\vartheta^\alpha= M \vartheta^\alpha$ does not
help, since this would change also the dimension of the Dirac field, in
order to retain  the physical dimension $[\hbar]$ of the Dirac action.

Thus the NY term $dC_{\rm TT}$ does NOT contribute to the {\em
chiral anomaly} in four dimensions, neither classically nor in QFT.
On would surmize that in  $n=2$ dimensional models only the term
$d\,^* {\cal A}$ survives in the heat kernel expansion, since it
then has the correct dimensions. However, it is well-known
\cite{PRs} that in 2D the axial torsion ${\cal A}$ vanishes
identically. Moreover, gravitational anomalies \cite{Leut},
specifically the Einstein anomaly and the Weyl anomaly, are fully
determined  by
 means of dispersion relations \cite{BK01}.

Let us stress the interrelation between the scale and chiral
invariance: The renormalized conformal (or trace) anomaly
\cite{DS93}
\begin{equation}
 \langle \vartheta^\alpha \wedge \sigma_\alpha\rangle=
 -{1\over
3\pi^2}\left[  Tr(G\wedge \,^*G)  +\frac{1}{24}\left(2R^{\alpha\beta\{\}}\wedge
 R^{\{\}\star}_{\alpha\beta} +d{\cal A}\wedge \,^*d{\cal A}\right)\right]
\label{confano}
\end{equation}
for the energy-momentum current $\sigma_\alpha:=\Sigma_\alpha-
D\mu_\alpha$ Belinfante symmetrized via the spin energy
(\ref{spinpot}) receives, in addition to the Riemannian Euler term,
a kinetic contribution of the Maxwell type from the axial torsion
${\cal A}$. The coefficients are similar to those in Eq.
(\ref{axano}), due to the fact that chiral and trace anomalies
constitute a supermultiplet \cite{Ya04}.

\section{Hamiltonian interpretation of anomalies}
In the canonical formulation \'a la Ashtekar \cite{Ash86}, the
translational NY term $dC_{\rm TT}$ plays via
\begin{equation}
 {\buildrel {(\pm)}\over{V}}_{\rm EC} := V_{\rm EC} \pm i d C_{\rm
TT} = \pm {1\over{2\ell^2}} Tr\, \left\{ (1 \mp \gamma_5)\,
\Omega\wedge\sigma\right\} = -{1\over{ 2\ell^{2}}} {\buildrel
{(\pm)}\over{R}}{}^{\alpha\beta}\wedge
\eta_{\alpha\beta}+{\Lambda\over{\ell^2}}\eta
\end{equation}
the role of the {\em generating functional} \cite{Mi92} for chiral,
i.e.  self-- or antiselfdual variables ${\buildrel
{(\pm)}\over{\Gamma}}$ in EC theory as well as in simple
supergravity \cite{mie86,MM99}.

The appearance of the Riemannian Pontrjagin term $dC_{\rm
RR}^{\{\}}$ in the anomaly (\ref{axano}) could pose  problems for
the canonical approach to gravity,
 since the
anomaly does not renormalize. In the presence of gravitational
instantons, which due to the necessary condition  $\Lambda \neq 0$
could even be the  dominating configurations, one gets a
net production of chiral zero modes and
a global symmetry is broken.

One could argue that this is a perturbative effect. In the Wilson
type loop approach to gravity \cite{Br92,Gr96,Mi02}, the tangential
complexified CS term $\underline{C}_{\rm RR}^{\{\}}$ is known
\cite{K90} to solve the Hamiltonian constraint $ {\cal H}_\Lambda \;
\Psi({\buildrel {(\pm)}\over{\underline{\Gamma}}}) =0 $ of gravity,
where the complex Ashtekar variable ${\buildrel
{(\pm)}\over{\underline{\Gamma}}}$ is the tangential part of the
self- or antiselfdual spin connection one--form. Since this solution
is intrinsically  non--perturbative, no anomaly should occur. In the
lattice gauge approach this is indeed the case, but the problem of
fermion doubling \cite{Sm86} appears to be another manifestation of
the anomaly.

It is instructive to look at the problem from an Hamiltonian point
of view since  the canonical formalism of chiral gravity is closely
related to the $SU(2)$ CS gauge theory on the three--dimensional
hypersurface, with
 ${\cal C} :=  \Gamma/{\cal G}$
of non-equivalent
gauge connections  as configuration space.

Gauge anomalies are  related to the global topology and have the
common feature  \cite{NA85} that the {\em Gauss constraint} ${\cal
G}^A \cong 0$ {\em cannot} anymore be implemented on the physical
states \cite{Fa84}. The reason is that the anomalous Ward identity
\begin{equation}
\L_n {\cal G}^A \cong n\rfloor(D\underline{\tau}^A)\, ,
\end{equation}
where $\L_n :=n\rfloor D +D n\rfloor$ is the gauge--covaiant Lie
derivative along the  normal direction, relates the time evolution
of the Gauss constraint to the  conservation law for the matter
current $\tau^A$ on the spacelike hypersurface \cite{Jiang}. Only
when the individual contributions to the anomaly cancel each other,
 a gauge theory can be consistently
quantized. In the EC formulation of the gravitationally coupled
Dirac field, it is  the canonical  spin $\tau^A:= (1/2)
\eta^{0A\beta\gamma}\tau_{\beta\gamma}$ which appears on the
right-hand side of the Gauss constraint. Since spin is via
(\ref{eq:spingamma},\ref{Dspin}) related to the axial current $j_5$,
it is precisely the {\em chiral anomaly} which prevents the Gauss
constraint to remain a proper  constraint under time evolution. This
result confronts the Ashtekar approach based on loop variables, and
thus on notions of parallel transport,
 with the chiral anomaly \cite{MK98,MK99}. The teleparallelism
 equivalent \cite{Mi92} of chiral gravity, where
 Wilson loops are replaced by Cartan circuits \cite{Mi99,Mi02}, may avoid
 some of these obstacles.

\section*{Appendix A: Gravitational Chern--Simons and Pontrjagin terms}

When the constant Dirac matrices $\gamma_\alpha$ obeying
$\gamma_\alpha\,\gamma_\beta
+\gamma_\beta\gamma_\alpha=2o_{\alpha\beta}$ are saturating the
index of the orthonomal coframe one--form
$\vartheta^\alpha=e_j{}^\alpha dx^j$ and its Hodge dual
$\eta^\alpha:={}^\ast\vartheta^\alpha$, we obtain a basis of
Clifford--algebra valued exterior forms \cite{mie86,Mi01} via:
\begin{equation}
\gamma:=\gamma_\alpha\vartheta^\alpha\,,\qquad
{}^\ast\gamma=\gamma^\alpha\eta_\alpha\,.\label{10-4.2}
\end{equation}

In terms of the Clifford algebra--valued {\em connection} $\Gamma :=
{i\over 4} \Gamma_i{}^{\alpha\beta}\,\sigma_{\alpha\beta} dx^i$, the
$SL(2,C)$--covariant exterior derivative is given by $D=d+
\Gamma\wedge$, where ${\sigma}_{\alpha\beta}= \frac{i}{2}
(\gamma_\alpha\gamma_\beta-\gamma_\beta\gamma_\alpha)$ are the
Lorentz generators entering  in the  two-form $\sigma:={i\over
2}\gamma\wedge\gamma = {1\over 2}\,{\sigma}_{\alpha\beta}
\,\vartheta^\alpha\wedge\vartheta^\beta$.

Differentiation of these basic
variables leads to the Clifford
algebra--valued {\em torsion} and {\em curvature} two--forms:
\begin{equation}
\Theta :=D\gamma =T^{\alpha}\gamma_{\alpha}\;, \qquad
\Omega := d\Gamma +\Gamma\wedge \Gamma =
{i\over 4}R^{\alpha\beta}\,\sigma_{\alpha\beta}
\label{tor}
\end{equation}
in RC geometry.  The {\em Chern--Simons term} for the
Lorentz connection  reads
\begin{equation}
C_{\rm RR}  := - Tr\, \left( {\Gamma}\wedge {\Omega} -
{1\over 3} {\Gamma}\wedge {\Gamma}\wedge  {\Gamma} \right)  \, .
\label{CRR}
\end{equation}
The corresponding {\em Pontrjagin topological term} can be obtained
by exterior differentiation
\begin{eqnarray}
dC_{\rm RR}  &=& - Tr\, \left( {\Omega}\wedge {\Omega}\right) \nonumber\\
&=& {1\over 2}R^{\{\}}_{\alpha\beta}\wedge R^{\{\}\alpha\beta} \nonumber\\
&+&\frac{1}{12}d\left[\,^*{\mathcal A}\wedge R^{\{\}}
-\frac{1}{3}{\mathcal A}\wedge d{\mathcal A} +\frac{1}{9}
\,^*{\mathcal A}\wedge^*({\mathcal A} \wedge  \,^*{\mathcal A})\right]
\label{Pontr}.
\end{eqnarray}
The latter contains \cite{MR06}, amongst others, a term proportional
to the curvature scalar $R:=\,^*(R^{\alpha\beta}\wedge
\eta_{\beta\alpha})$ and  the axial torsion piece $d{\cal A}\wedge
d{\cal A}$ of the axial anomaly with a relative factor 9 as required
by the supersymmetric path integral \cite{Ma88}.

Since the coframe is the `soldered'  translational part
\cite{MMNH,TM00} of the Cartan connection, a related {\em
translational} CS term arises
\begin{equation}
C_{\rm TT}  :=  {1\over{8\ell^2}} Tr\, ( {\gamma} \wedge  {\Theta} )=
{1\over{2\ell^2}}\, {\vartheta^\alpha}\wedge T_{\alpha}\,.
\label{CTT}
\end{equation}
 By exterior differentiation we obtain the NY four--form \cite{NY}:
\begin{equation}
dC_{\rm TT}
= {1\over{8\ell^2}} Tr\, (\Theta\wedge \Theta -4i\Omega\wedge\sigma)
={1\over{2\ell^2}} \left(T^\alpha\wedge
T_\alpha+R_{\alpha\beta}\wedge\vartheta^\alpha\wedge\vartheta^\beta\right) \,.
\label{eq:NY}
\end{equation}

It is crucial to note that a fundamental length $\ell$ necessarily occurs here for
dimensional reasons. This can be also understood
by a de Sitter type  gauge approach, in which the $sl(5,R)$--valued
 connection   $\hat\Gamma =\Gamma +(\vartheta^\alpha L^4{}_\alpha +
 \vartheta_\beta L^\beta{}_4{})/\ell$ is expanded into the dimensionless linear connection
 $\Gamma$ plus the coframe $\vartheta^\alpha= e_i{}^\alpha\, dx^i$
which carries canonical dimension $[length]$.
The corresponding
Pontrjagin term $\hat C_{\rm RR}$
splits via
\begin{equation}
\hat C_{\rm RR} =C_{\rm RR} -2 C_{\rm TT} \label{Poin}
\end{equation}
into the linear one and the translational CS term, see the footnote
31 of Ref. \cite{PRs} for details.

\begin{theacknowledgments}
 We would like to thank  Hugo Morales--T\'ecotl and Luis Urrutia
for useful hints and comments. This article is dedicated to the
memory of Yuval Ne'eman.
\end{theacknowledgments}

\end{document}